\documentclass[doublecol]{epl2}

\usepackage{graphicx}
\usepackage{amsmath}
\usepackage{amsfonts}
\usepackage{amssymb}
\usepackage{psfrag}

\newcommand{\mean}[1]{\left\langle #1 \right\rangle}
\newcommand{\island}{island}
\newcommand{\islands}{\island{}s{}}

\newcommand{\ltun}{\lambda}			
\newcommand{\dfric}{\gamma}			

\newcommand{\eog}{\eta/\dfric}		

\begin{document}

\author{
Alexander Croy\inst{1,2}\footnote{Email: alexander.croy@chalmers.se} \and Alexander Eisfeld\inst{1,3}\footnote{Email: eisfeld@pks.mpg.de}
}
\shortauthor{A. Croy and A. Eisfeld}

\institute{
	\inst{1} Max-Planck-Institute for the Physics of Complex Systems,
 N\"{o}thnitzer Str.~38, 01187 Dresden, Germany\\
	\inst{2} Department of Applied Physics,
Chalmers University of Technology, 41296 G\"{o}teborg, Sweden\\
	\inst{3} Department of Chemistry and Chemical Biology,
Harvard University, 12 Oxford Street, Cambridge, MA 02138, USA
}

\title
{Dynamics of a nano-scale rotor driven by single-electron tunneling}

\date{\today}


\abstract{
We investigate theoretically the dynamics and the charge transport properties of a rod-shaped nano-scale rotor, which is driven by a similar mechanism as the {\it nanomechanical single-electron transistor (NEMSET)}.
We show that a static electric potential gradient can lead to self-excitation of oscillatory or continuous rotational motion.
We identify the relevant parameters of the device and study the dependence of the dynamics on these parameters.
We discuss how the dynamics is related to the measured current through the device.
Notably, in the oscillatory regime, we find a negative differential conductance. 
The current-voltage characteristics can be used to infer details of the surrounding environment which is responsible for damping.
}

\maketitle

\section{Introduction}
In recent years it has emerged that the coupling of electrical and mechanical degrees of freedom on the nanometer scale provides the opportunity to build novel devices, extending the concepts of conventional electronic devices \cite{cr00,bl04}. 
A seminal example is the nanomechanical single-electron transistor (NEMSET) \cite{GoIsVo98_4526_,IsGoVo98_150_,ShGaGo03_R441_,MoGoKo09_241403_}, where electrons can tunnel from a source to a drain electrode via a movable island or grain, whose charge (i.e.\ the number of electrons that occupy the island) is determined by the Coulomb-blockade effect. 
Since the tunneling amplitude depends exponentially on the position of the island, the current is very sensitive to the mechanical motion. 
 For a sufficiently large bias voltage a self-excitation of periodic oscillations occurs in conjunction with charging and de-charging of the island. This mechanically assisted charge transport is called electron shuttling \cite{GoIsVo98_4526_}. 
The grain is embedded in a medium which creates a restoring-force and determines essentially the eigenfrequency of the oscillation of the shuttle.

In this letter we consider a device which can rotate freely and which is driven by the same mechanism as the electron shuttle described above. The driving force is determined by a static voltage, which is applied along the device \cite{WaVuKr08_186808_,SmSaMo08_031921_,smmo+09}.
In contrast to the conventional shuttles this results in a force that depends non-linearly on the relevant system coordinate.
The coupling of mechanical motion and tunneling leads, as we will show, to the self-excitation of oscillatory and rotational motion even in the presence of damping. 
The frequency of oscillation (rotation) depends on the ratio of the driving force and the friction.

While being based on the same principle as the charge shuttle, the present device exhibits markable differences.
In the oscillatory regime, the current through the device decreases with increasing bias voltage, which is surprising from the conventional shuttling point of view. We
attribute this effect to the presence of a separatrix in the phase space, which separates
oscillations from rotations. Approaching the separatrix involves a slowing down of
the dynamics and decreasing oscillation amplitudes, which results in the decreasing current.
In the rotational regime, one may realize a nano-scale motor, which is
driven by a static voltage \cite{smmo+09}.

Using numerical calculations supplemented by an analytical analysis, we find that the dynamics of the rotor is governed by three dimensionless parameters: tunneling length, field strength and damping constant. This circumstance allows us to predict, for example, a transition from oscillatory to rotational motion fairly independent of the actual realization of the device. Our analysis provides an intuitive picture of the dynamics for a large range of parameters.

\section{Model}
\begin{figure}[bt]
	\centering
   \includegraphics[width=6cm]{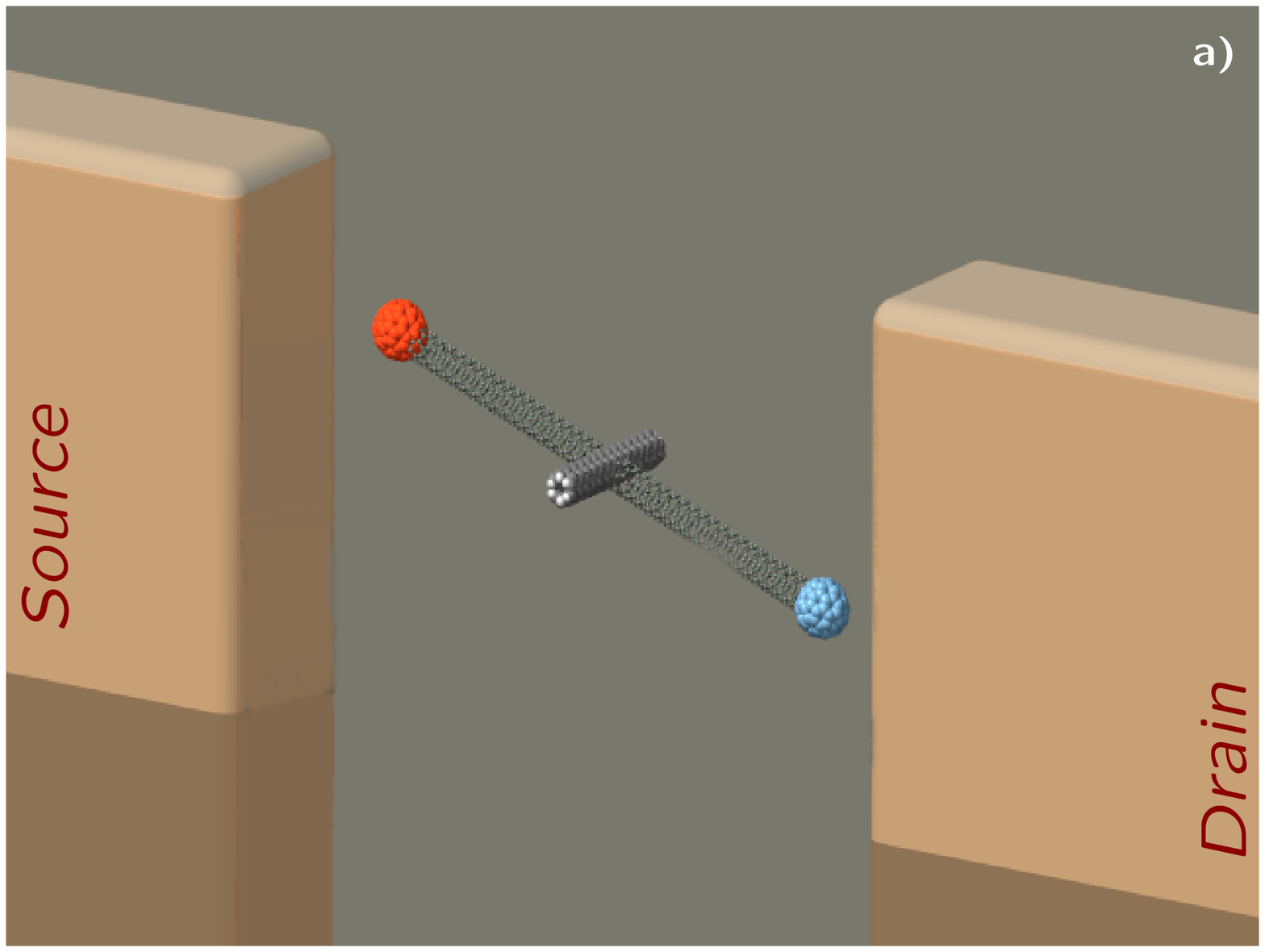}\\
	\includegraphics[width=6cm]{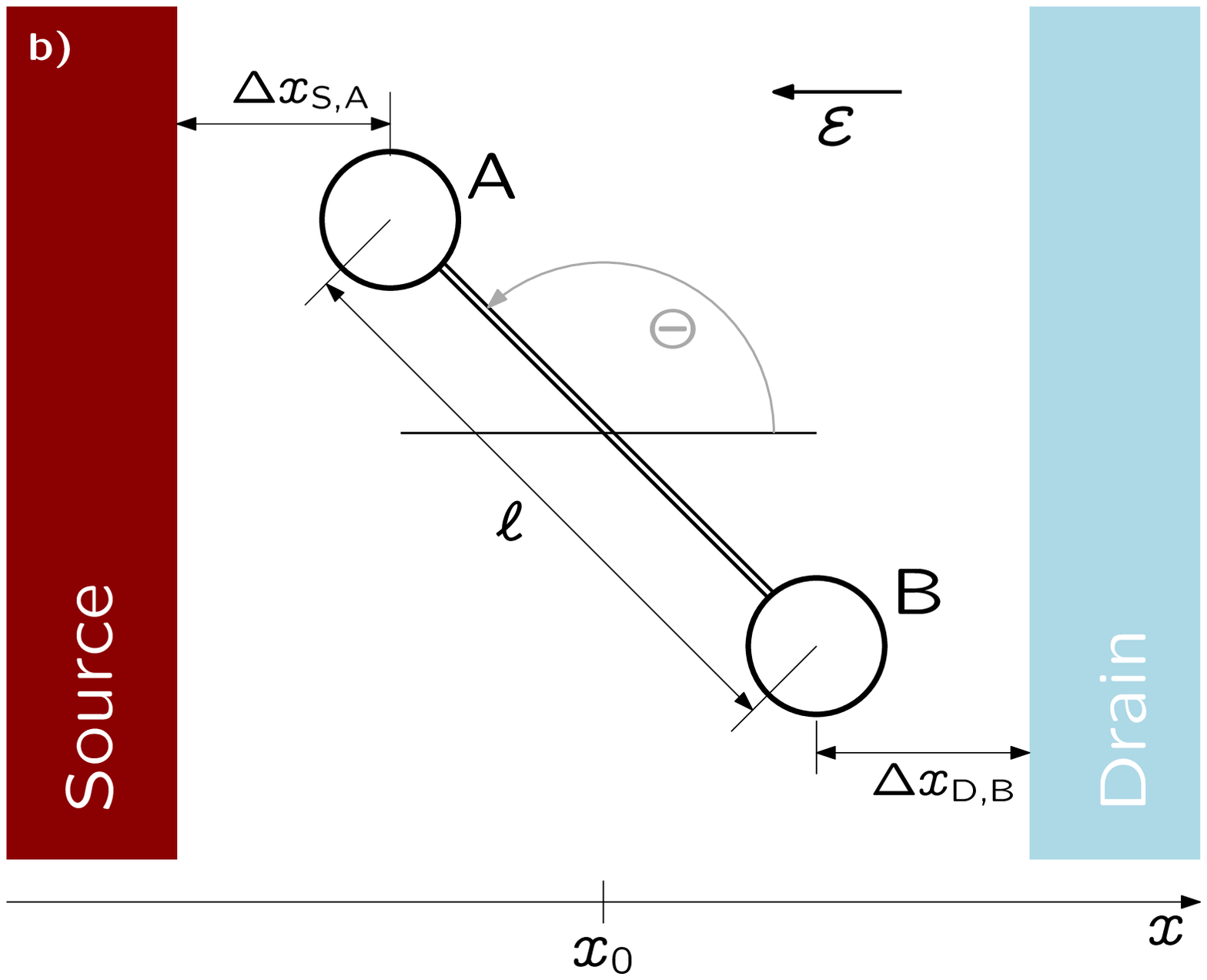}
 	\caption{(Color on-line) a) Artists view of a possible realization of the single electron motor using
 		(multi-wall) carbon nanotubes and two charge \island{}s. 
 		b) Sketch of the relevant parts and coordinates.
	}
 	\label{fig:example}
\end{figure}
A possible realization of the device, discussed in the present letter, is sketched in fig.~\ref{fig:example}.
It consists of two charge \island{}s (e.g.\ quantum dots, metallic nano-clusters, etc.), which we denote by A and B, placed at the ends of a rigid rod of length $\ell$.
This structure can rotate in the x-y plane around an axis in the middle of the rod. 
The support shaft of the rotor could be realized similar to the one of ref.~\cite{FeYuHa03_408_} by multi-walled carbon nanotubes or in a way proposed in ref.~\cite{WaVuKr08_186808_}.
The motion of the rotor around the pivot axis is characterized by the angle $\Theta$ and the corresponding moment of inertia $I$.

To drive the device, the rotor is placed between source (S) and drain (D) contacts which are connected to electron reservoirs. The contacts are kept at different chemical potentials by an externally applied bias voltage.
The electrons may tunnel between the contacts and the charge \island{}s. 
The respective tunneling amplitudes depend sensitively on the position of the \island{}s.
Specifically, we assume an exponential dependence \cite{GoIsVo98_4526_}. 
For example, the tunneling amplitude from the source to \island{} A is
$
 	T_{\rm S,A} \propto \exp[ -\Delta x_{\rm S,A}/\ltun_{\rm S,A} ]
$,
where $ \Delta x_{\rm S,A}$ is the distance between the source contact and \island{} A 
and $\ltun_{\rm S,A}$ is the tunneling length.
Similar expressions are used for the tunneling amplitudes $T_{\rm D,A}$,
$T_{\rm S,B}$ and $T_{\rm D,B}$.
It is convenient to introduce the dimensionless ratio
\begin{equation}
	\xi_{\rm S,A} = \frac{\ell}{\ltun_{\rm S,A}}\;,
\end{equation}
which can be regarded as a measure of the change of distance with respect to the tunneling length.

\begin{figure*}[tb]
    \centering
\psfrag{LL}[t][t][1][270]{$\Lambda$}
    \includegraphics[width=0.8\textwidth]{Fig2ab-phase-ovrvw}
	 \hspace{0.4cm}
    \includegraphics[width=0.12\textwidth]{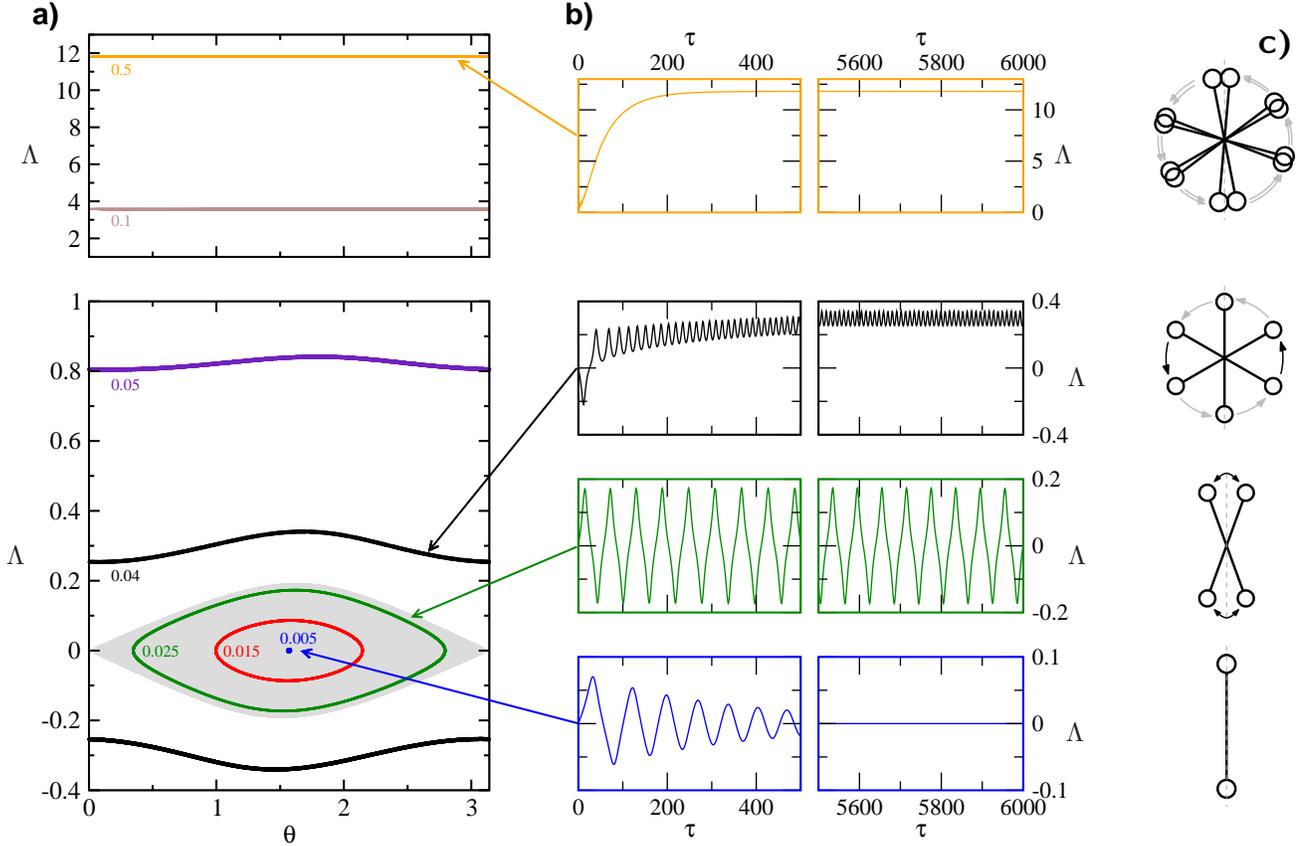}
    \caption{(Color online) Dynamics of the rotor for various 
    $\eta$ and fixed $\dfric = 0.01$ and $\xi=2$.
(a) Phase portrait of the steady-state dynamics. Within the gray-shaded area an oscillatory motion takes place. The values of $\eta$ are given next to the respective trajectory.
(b) Exemplary time evolution that leads to the corresponding steady-state dynamics. 
(c) Sketch of the dynamics for the four cases shown in b).
    }\label{fig:phase}  
\end{figure*}

Since the \island{}s are nanoscale objects, their charging will be determined by the Coulomb-blockade effect \cite{avko+91,grde92}. For sufficiently small tunnel coupling electrons are transferred sequentially. According to the orthodox theory of Coulomb blockade \cite{avko+91,be91,mewi+91} the tunneling rates are (for \island{} A)
\begin{subequations}\label{eq:tunnrates}
\begin{align}
	W_{\rm S,A}(\Theta)=& \Gamma_{\rm S A} {\rm e}^{-\xi_{\rm S,A}\cos(\Theta)}\;,\\
	W_{\rm D,A}(\Theta)=& \Gamma_{\rm D A} {\rm e}^{+\xi_{\rm D,A}\cos(\Theta)}\;,
\end{align} 
\end{subequations}
where for example $\Gamma_{\rm S A}$ is defined as the tunneling rate from source to dot A when the rotor is in the perpendicular position, i.e., for $\Theta=\pi/2$. 
Similar relations hold for the tunneling rates of \island{} B (replacing $\cos(\Theta)$ by $\cos(\Theta+\pi)$).

In the following, we consider a very large charging energy such that only 
one excess electron can occupy each \island{} at the same time (Coulomb blockade regime). 
Further, for simplicity, we assume that the bias voltage is very large (large bias limit).
The large bias approximation holds as long as the respective broadened energy level is completely in the transport window, i.e., for source-drain voltages 
$e V_{\rm SD} \gg k_{\rm B}T$ and $e V_{\rm SD} \gg\Gamma_{\rm S A/B} + \Gamma_{\rm D A/B}$. Here, $k_{\rm B}$ is the Boltzmann constant and $T$ denotes the temperature of the electronic
reservoirs.
Then, electron transport occurs only in one direction \cite{gupr96}:
From the source contact electrons can tunnel onto the \island{}s.
On the other side they can tunnel from the \island{}s to the drain contact.

\section{Dynamics}
Denoting the electronic population of island A (B) by $P_{\rm A (B)}$ we can write for average charges of the \island{}s $q_{\rm A (B)}= -e P_{\rm A (B)}$ where $ P_{\rm A (B)}$ can attain values between $0$ and $1$.
Due to the tunnel coupling to the leads the populations $P_{\rm A (B)}$ are time-dependent and determined by the rate equation
  \begin{equation}\label{eq:pop}
      \dot{P}_{\rm A} 	= W_{\rm S,A}(\Theta) \left( 1 - P_{\rm A} \right)-W_{D,A}(\Theta)  P_{\rm A} \;.\
  \end{equation}
The first term on the right hand side describes tunneling from the source onto \island{} A and the second term is responsible for tunneling from the \island{} to the drain contact.  

When \island{} A possesses a charge $q_{\rm A}$ then it experiences a force $F_x$ in $x$-direction which is proportional to the electric field strength $\mathcal{E}\propto V_{\rm SD}$ 
induced by the source-drain voltage $V_{\rm SD}$ between the contacts: 
$
 F_x= q_{\rm A}\, \mathcal{E}
$.
Then, in the mean field description introduced above, the torque acting on the rotor is given by
$
 M_{\rm A}= -e\, P_{\rm A}\, \mathcal{E}\, \frac{\ell}{2}\, \sin(\Theta)
$.
A similar expression holds for \island{} B (replacing $\sin(\Theta)$ by $\sin(\Theta+\pi)$).

To keep the discussion transparent, in the following we will take all tunneling rates to be equal\footnote{
In practice the couplings of the different \island{}s to the two leads will not be perfectly identical. From our numerical simulations we have found that the features discussed in the present work are quite robust with respect to changes in the relative couplings. For example, even for very asymmetric tunneling rates $\Gamma_{A,S}/\Gamma_{A,D}\approx 4$ the results presented in the following, remain qualitatively the same.} and denote them by $\Gamma$. This rate can then be used to introduce a dimensionless time,
\begin{equation}
  \label{eq:tau}
	\tau = \Gamma\, t\;,
\end{equation}
and a dimensionless field strength,
\begin{equation}
	\eta \equiv e \mathcal{E} \ell/(2 I \Gamma^2)\;.
\end{equation}
Due to the torque $M_{\rm A} + M_{\rm B}$ there will be a change in angular momentum $L$ of the rotor.
For the dimensionless angular momentum
\begin{equation}
	\Lambda = \frac{L}{I \Gamma}
\end{equation}
the equation of motion reads
\begin{equation}\label{eq:class_reduced_L}
\begin{split}
\frac{\partial }{\partial \tau}\Lambda &= - \eta \left( \sin(\Theta) P_{\rm A} + \sin(\Theta + \pi) P_{\rm B} \right) -F(\Lambda) \\
&= -\eta \sin(\Theta)\Delta P-F(\Lambda)  \;,
\end{split}
\end{equation}
where we have added a phenomenological damping term $F(\Lambda)$ and in the last line we have defined $\Delta P=P_{\rm A}-P_{\rm B}$. 
In the following calculations we choose the damping to be linearly dependent on the velocity \cite{co94, WaVuKr08_186808_}, i.e., $F(\Lambda)=\dfric \Lambda$, with a dimensionless damping constant $\dfric$.
The damping will typically be accompanied by fluctuations, whose influence will be discussed in a later section.

The population difference $\Delta P$, which appears in eq.\ \eqref{eq:class_reduced_L}, also depends on time. We have
\begin{equation}
\label{eq:Delta_P/dt}
\frac{d \Delta P}{d\tau}= - 2 \sinh[\xi \cos(\Theta)]
-2 \cosh[\xi \cos(\Theta)]\Delta P \;,
\end{equation}
where we have also set all tunneling lengths equal to $\xi$.

Finally, the angle $\Theta$ is related to the momentum $\Lambda$ by
\begin{align}
	\label{eq:class_reduced_Th}
	\frac{\partial \Theta}{\partial \tau} &= \Lambda \;.
\end{align}
The three coupled equations, \eqref{eq:class_reduced_L}, \eqref{eq:Delta_P/dt} and \eqref{eq:class_reduced_Th}, govern the dynamics of the system. 
In the following, we will investigate how this dynamics depends on the parameters $\xi$, $\dfric$ and $\eta$.
We are mainly interested in the steady-state dynamics at long times after the initial irregularities have vanished.

For given damping $\dfric$ and tunneling length $\xi$ we find three qualitative different regimes of the steady-state motion depending on the values of the driving $\eta$. These are illustrated in fig.\ \ref{fig:phase}{c}.
In fig.\ \ref{fig:phase}a,b numerical solutions of the set of equations \eqref{eq:class_reduced_L}-\eqref{eq:class_reduced_Th} are shown for $\dfric=0.01$ and $\xi=2$.

In fig.~\ref{fig:phase}{a} the steady-state solutions for various values of $\eta$ are depicted by representing the angular momentum $\Lambda$ as a function of the angle $\Theta$. Different initial conditions $\Theta_0$ lead to the same
steady-state solution up to the direction of rotation, which is opposite for $\Theta_0$
and $-\Theta_0$.
In fig.\ \ref{fig:phase}b examples of the initial time dependence of the angular momentum is shown for a situation where the rotor is initially at rest at an angle $\Theta=0.4$. 

\begin{figure}[tbp]
	\centering
	 \includegraphics[width=0.4\textwidth]{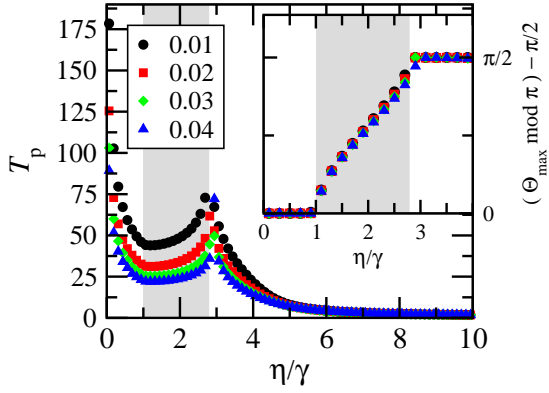}
  \caption{\label{fig:periode}(Color online) Period of steady-state oscillations/rotations is shown as a function of $\eta/\dfric$. Different symbols (colors) indicate different values of $\dfric$, as given in the legend. The inset shows the corresponding maximal angle relative to $\Theta=\pi/2$. 
  } 
\end{figure}

For $\eta<\gamma$ the rotor shows a relaxation towards the equilibrium position given by $\Theta=\pi/2$, i.e., both \islands{} A and B are in the middle between the source and the drain contact. 
Obviously, in this regime the dynamics of the rotor is dominated by the damping, which leads to the motion of the rotor being slow compared to the tunneling time. Consequently, the populations are in equilibrium with the reservoirs at each instant of time. Additionally, the strong damping leads to decreasing amplitudes and, in the symmetric case considered here, the population difference vanishes accordingly. 
Since the rotor is not accelerated anymore, the motion eventually grinds to a halt.
Electrons will tunnel with the constant rate $\Gamma$ to/from the \islands{} to the contacts. The stationary solution is given by the single (blue) point in fig.~\ref{fig:phase}a, located at $\Theta=\pi/2$ and $\Lambda=0$. 

Upon increasing $\eta$ one reaches a regime where the rotor oscillates around the position 
$\Theta=\pi/2$ and the force on the charged \islands{} is not yet strong enough to lead to a rotational motion. This regime is characterized by closed curves within the shaded area in fig.~\ref{fig:phase}a. The outmost trajectory of this area was obtained from an adiabatic approximation to eq.~\eqref{eq:Delta_P/dt} and considering extremal swing.
The occurrence of the oscillatory regime is a direct consequence of the distance dependence in the tunneling rates. For sufficiently weak damping, inertia leads to
slightly increasing amplitudes in each period. Due to the increasing tunneling rates, this comes along with an increase of the population difference and therefore a larger backward acceleration. Of course the increase of the oscillation amplitude will be
limited by the damping, which becomes more and more important as the
velocity (or amplitude) increases. Consequently, one has two competing
mechanisms: one that favors to increase the oscillation amplitude
(electric field/tunneling) and one that acts to decrease it (damping). The result is
the occurrence of stable oscillations in the long-time limit. Notice, that these oscillations are akin to the self-excited shuttling in NEMSET \cite{GoIsVo98_4526_}.
However, in contrast to the conventional charge shuttle, for the rotor the period of the oscillation increases with increasing $\eta$.
This behavior is illustrated in fig.~\ref{fig:periode}, where the period of steady-state oscillation/rotation is shown as a function of $\eta/\dfric$.
In the oscillatory regime, indicated by the gray-shaded area, a clear increase of the period of oscillation can be seen, spiking at $\eta/\gamma \approx 2.8$. 
This point corresponds to the separatrix as shown in fig.\ \ref{fig:phase}a.

For even larger values of $\eta$ periodic rotational motion sets in (see the curves for $\eta=0.04$ and $\eta=0.05$ in fig.~\ref{fig:phase}a). The direction of rotation in the long-time limit depends on the initial conditions. In this regime, the motion is dominated by the driving through the electric field. The rotational energy becomes large enough to overcome the damping and to allow an almost free rotation, characterized by a constant angular momentum. This case is shown in fig.~\ref{fig:phase}a in the upper panel.

\section{Angular momentum} With respect to a possible application as a motor, the temporally averaged angular momentum in the steady-state is an important quantity.
Thus, in our numerical simulations we calculate 
\begin{equation}
	\label{eq:l_aver}
	\langle \Lambda \rangle = \frac{1}{\mathcal{T}}\int_{t_1}^{t_1+\mathcal{T}}\Lambda(t) \rm{d} t \;,
\end{equation} 
where we have chosen the time $t_1$ large enough to ensure that we are in the regime of stationary dynamics and we have taken the time interval $\mathcal{T}$ to cover many periods of oscillation (rotation).
In fig.~\ref{fig:Averages} numerical results for various $\xi$ and $\dfric$ are shown. The different colors of the dots represent different values of $\dfric$ for the same $\xi$ as in fig.~\ref{fig:phase}. As can be anticipated from eq.\ \eqref{eq:class_reduced_L}, with $F(\Lambda)=\dfric \Lambda$, the curves for different values of $\dfric$ nearly coincide.

In the upper row of fig.~\ref{fig:Averages} the dependence of $|\langle \Lambda \rangle|$ on $\eog$ is shown.
Obviously, in the damping dominated regime the average angular momentum is zero, since
$\Lambda\to0$ in the stationary limit. Also in the oscillatory regime (indicated by the gray-shaded area) one has $\langle \Lambda \rangle=0$, since the angular momentum changes its sign in each period (see, e.g., the third row of fig.~\ref{fig:phase}b). 
Thus, the averaged $\Lambda$ is not an ideal quantity to gain insight into the oscillatory regime.
However, it is well suited in the rotational regime, which is characterized by a finite value of the average angular momentum $\langle \Lambda \rangle$ and its modulus\footnote{When considering $\langle\Lambda\rangle$ instead of $|\langle\Lambda\rangle|$ a branching of the curve appears, where the upper and lower branches describe anti-clockwise or clockwise rotation, respectively. The direction of rotation depends in a non-trivial way on the initial angle $\Theta_0$. Due to the symmetry of system, changing $\Theta_0$ to $-\Theta_0$ reverts the direction.}.
In the left column of fig.~\ref{fig:Averages} the focus is on relatively small values $\eog < 10$, in the right column the region from $\eog=10$ to $\eog=150$ is covered. 
In this region the dependence of the angular momentum on $\eog$ can be estimated analytically:
In the stationary limit, the time-averages of the dissipated work and
the work due to accelerating the charges have to be equal,
$ 
 		\mean{W_{\rm fric}} = \mean{W_{\rm charge}}
$. Changing the time-average to an angular average amounts to
$
	\dfric \int^{2\pi}_0 d\Theta\, \Lambda(\Theta) = 
		-\eta \int^{2\pi}_0 d\Theta\, \Delta P(\Theta) \sin\left( \Theta \right)
$.
For large values of $\eog$, the angular momentum is constant, $\mean{ \Lambda}\approx \Lambda\equiv \Lambda_{\rm stat}$. After integration by parts and replacing the resulting derivative of the population difference by eq.~\eqref{eq:Delta_P/dt}, $d\Delta P/d\Theta = (dP/d\tau)/(d\Theta/d\tau)$, one obtains 
\begin{equation}\label{eq:lambdaavgapprox}
	\Lambda_{\rm stat} \approx \sqrt{c(\xi) \frac{\eta}{\dfric}-d(\xi)}
\end{equation}
with 
$
c(\xi)=\mean{ \cos\left( \Theta \right)\ 2 \sinh\left( \xi \cos(\Theta)\right) }_\Theta
$
and 
$
d(\xi)=\mean{\cos\left( \Theta \right)\ 2 \cosh\left( \xi \cos(\Theta)\right) \Delta P}_\Theta
$.
Using the Jacobi-Anger identity, the first term becomes
$
c(\xi) =2 I_1(\xi),
$
where $I_n$ is the $n$th order modified Bessel function of the first kind.
The second term is more difficult to evaluate. To eliminate the population difference, one 
has to repeatedly use eq.\ \eqref{eq:Delta_P/dt}. However, in the limit $\eog\to\infty$ 
this term vanishes.

We have found that these analytical estimates describe our numerical results very well in the region of large $\eog$. 
For the case $\xi=2$, shown in figs.~\ref{fig:phase} and \ref{fig:Averages} we have numerically found $c(\xi)\approx 3.18$ and $d(\xi) \approx 19.76$.
We found also very good agreement for other values of $\xi$.

\section{Current}
Since the dynamics of the rotor is essentially driven by charge transport it is instructive to take a closer look at the electric current.
The time-dependent current (in units of the electron charge $e$) is given by 
\begin{equation}
	\label{eq:current(t)}
	J(t)/e=W_{\rm A,D}\big(\Theta(t)\big)\, P_A(t)+W_{\rm B,D}\big(\Theta(t)\big)\, P_B(t)\;.
\end{equation}
As before we are primarily interested in the time-averaged current $\langle{J}\rangle$, which we define in the same manner as in eq.~(\ref{eq:l_aver}).
Note, that the resulting stationary current is independent of the initial condition and, in particular, of the direction of rotation.  

\begin{figure}[tb]
	\centering
	\includegraphics[clip,width=\columnwidth]{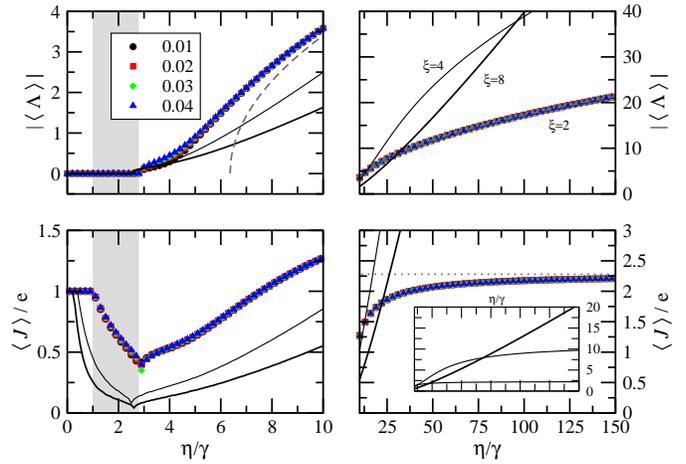}
	\caption{\label{fig:Averages} (Color online) Dependence of the average angular momentum $\langle \Lambda \rangle$ and the average current $\langle J \rangle$ on the ratio $\eog$. Symbols denote results for $\xi=2$ and different values for $\dfric$ as given in the legend.
Right column: focus on large values of $\eog$. Left column: details for small $\eog$. The dashed line indicates the behavior obtained from eq.~(\ref{eq:lambdaavgapprox}) and the dotted line shows the asymptotic value of the stationary current 
$\langle J \rangle_{\rm stat}=\Gamma\, I_0(\xi)$. The inset shows $\langle J \rangle$ for the same $\eog$, but on a larger scale.
	}
\end{figure}

Numerical results for $\xi=2$ are shown in the second row of fig.~\ref{fig:Averages}.
Here the average current is plotted as a function of $\eog$ for different $\dfric$ (the same as for the angular momentum in the upper row).
Again, all curves coincide quite well.
From the behavior of the averaged current one can clearly distinguish the regime where the rotor oscillates back and forth (corresponding to the shaded area) and the regime corresponding to a motionless rotor ($\eog < 1$).

For a very small ratio $\eog$ the current is constant, since the rotor does not move and electron transport takes place by tunneling to and from the \island{}s in the middle between the contacts. In contrast, in the region where the rotor oscillates one finds a rapid decrease of the average current. When the rotational motion starts the current rises again slowly. 
The decrease in the current is related to the fact, that the mechanical motion is becoming slower as the separatrix, which separates the oscillation trajectories from the rotating ones, is approached (cf.\ fig.\ \ref{fig:periode}). 
With increasing $\eta/\dfric$ the amplitude of the oscillations increases and the rotor spends an increasing time at the reversal point, where the tunneling to the opposite electrode is strongly diminished. This results in a decrease of the current.

In the limit of a very large ratio $\eog$, the stationary current attains a constant value, which is readily calculated from eqs.\ \eqref{eq:pop} and \eqref{eq:current(t)} by replacing the rates by their time averages. One obtains
$\mean{ W_{\rm S,A} }= \Gamma_{\rm S A} I_0(\xi_{\rm S,A})$
and
$\mean{ W_{\rm D,A} }= \Gamma_{\rm D A} I_0(\xi_{\rm D,A})$,
where $I_0(\xi)$ denotes the modified Bessel function of the first kind.
The stationary current (for the case of identical tunnel parameters) is then given by $\langle J \rangle_{\rm stat}=\Gamma\, I_0(\xi)$.
This value is depicted in fig.~\ref{fig:Averages} as the horizontal dotted line.
It is in very good agreement with the numerical results.

In addition to the case $\xi=2$, in fig.~\ref{fig:Averages} also numerical results for other values of $\xi$ are shown.
One sees that for larger $\xi$ the drop in the current sets in for smaller values of $\eog$ and is more pronounced.

\section{Robustness of the results}
In the following, we would like to briefly comment on the robustness of our results with respect to asymmetries in the couplings to the leads and fluctuations of the environment (e.g.\ related to internal friction or interaction with some background gas at finite temperature).
In the present work we have restricted ourselves to an ideal situation, neglecting these effects, since we wanted to keep our analysis as clear as possible. 
Our results are only weakly affected by including
different couplings and fluctuations.
For the couplings this has been discussed in connection with eq.~(\ref{eq:tau}).
To gain insight into the influence of fluctuations we included a Langevin term in eq.~(\ref{eq:class_reduced_L}) and solved the resulting stochastic differential equation numerically.
For small $\eta$ the fluctuations have the largest effect. 
In the oscillatory regime and close to the separatrix they can lead to sporadic rotations. Similarly, 
in the standstill regime oscillations with a few periods can be triggered, resulting in a small decrease of the current. Overall, 
this leads to a smoothing of the observed features for the time-averaged quantities.
However, for moderate fluctuation strengths (approximately smaller than the dissipative energy $\gamma \langle \Lambda \rangle$) the qualitative behavior is not changed. 
In the rotational regime (large $\eta$) our results were (as expected) nearly unaffected by fluctuations.

\section{Conclusions}
We have investigated the dynamics 
of a nanoscale rotor which is driven 
by the sequential tunneling of electrons between electronic contacts and the device. 
The interplay of the electric and the mechanical degrees of freedom leads to three different regimes for the motion of the rotor, which can be obtained by adjusting the external bias voltage: stationary, oscillatory and rotation. 

The oscillatory regime has some similarities with the oscillatory motion of the conventional charge shuttles.
However, in contrast to the charge shuttles, the motion of the rotator is slowed down with increasing driving strength.
This behavior results in a {\it decrease} of the electric current with {\it increasing} bias voltage. In the rotational regime the current starts to increase again
and the device can be considered as a nano-scale motor.
It is worth noting that the setup discussed in the present paper has some similarities with bio-motors \cite{SmSaMo08_031921_}, such as the F0 motor of adenosine triphosphate synthase.
Further examples of molecular rotors can be found in ref.~\cite{KoClHo05_1281_}.

One might speculate about possible applications of the presented device.
Due to the high sensitivity of the current 
to the ratio $\eog$, the device may be used to detect changes in its environment.
For example, the damping constant $\dfric$ may be determined by tuning the
voltage and maintaining the same current. In this scenario the rotor can be used as
a sensor.
Another possible application is given in the context of electron pumping. Especially in the rotation regime the charge transport is determined 
by the mechanical motion. This provides the possibility of a mechanically
stabilized pumping in the high frequency regime.
In view of recent achievements regarding the cooling of nanomechanical systems to the quantum regime \cite{ocho+10,tedo+11,poza11,chma+11}, the questions if and how it is possible to cool a nanomechanical rotor and its interplay with quantum gears \cite{ma02} will be an interesting subject for further studies.

Altogether, the investigated nano-scale rotor is an interesting nanoelectromechanical device with potential applications as a sensor or a single-electron motor.

\begin{acknowledgments}
The authors would like to thank Prof.\ Saalmann for his helpful
comments during the preparation of the manuscript.
\end{acknowledgments}


\end{document}